\documentclass[11pt]{article}
\usepackage{amsmath, amsthm}
\usepackage{latexsym}
\usepackage{graphicx}
\makeatletter
\newfont{\footsc}{cmcsc10 at 8truept}
\newfont{\footbf}{cmbx10 at 8truept}
\newfont{\footrm}{cmr10 at 10truept}
\makeatother
\newtheorem{proposition}{\bf Proposition}

\begin{document}
\title{Efficient Simulation of a Bivariate Exponential Conditionals Distribution}

\author{Yaming Yu\\
\small Department of Statistics\\[-0.8ex]
\small University of California\\[-0.8ex] 
\small Irvine, CA 92697, USA\\[-0.8ex]
\small \texttt{yamingy@uci.edu}}

\date{}
\maketitle

\begin{abstract}
The bivariate distribution with exponential conditionals (BEC) is introduced by Arnold and Strauss [Bivariate
distributions with exponential conditionals, J. Amer. Statist. Assoc. 83 (1988) 522--527].  This work presents a simple and
fast algorithm for simulating random variates from this density.

{\bf Keywords:} Bivariate exponential conditionals; rejection algorithm; simulation.
\end{abstract}

\section{Introduction}
Arnold and Strauss (1988) introduced a bivariate distribution with exponential conditionals (BEC), whose unnormalized 
probability density function is specified by
\begin{equation}
f(x, y)= e^{-(\beta x+\gamma y + \delta\beta\gamma xy)},\quad x>0,\quad y>0,
\label{dens}
\end{equation}
for $\beta>0,\ \gamma>0,\ \delta\geq 0$.  Distribution theory, methods of estimation, and related specifications of joint 
distributions by conditionals can be found in Arnold and Strauss (1988) and Arnold and Strauss (1991).  The BEC distribution 
in particular has received attention in applications such as reliability analysis (Nadarajah and Kotz, 2006).

This paper is concerned with simulating random variates from the BEC densities.  Arnold and Strauss (1988) actually suggested 
a rejection method using a product exponential as the proposal density.  While this method is convenient and efficient 
for some parameter configurations (specifically, when $\delta$ in (\ref{dens}) is small), it can be shown that as 
$\delta\rightarrow \infty$, its acceptance rate approaches zero.  To design more efficient algorithms, we may explore several 
general approaches (Devroye, 1986), e.g., inversion, rejection, and ratio of uniforms.  Direct inversion is difficult in this 
case because the inverse distribution function of either $X$ or $Y$ is not readily available, we therefore consider a 
rejection method.  The ratio of uniforms method is considered, but the resulting algorithm is not presented here because it 
also deteriorates as $\delta\rightarrow \infty$ and is clearly inferior to the proposed method. 

Section 2 presents the new rejection method and evaluates its performance.  With a careful choice of the envelope function, 
we obtain an acceptance rate of at least 70\%, while keeping the algorithm simple and easy to implement.  

\section{A New Rejection Algorithm}
A convenient rejection method, suggested by Arnold and Strauss (1988), is to sample $(X, Y)$ from the (unnormalized) proposal 
density, or envelope function
$$g_0(x, y)=e^{-\beta x-\gamma y},\quad x>0,\quad y>0,$$
and then accept $(X, Y)$ with probability $f(X, Y)/g_0(X, Y)=e^{-\delta\beta\gamma XY}$.  This may be implemented as the 
following Algorithm A.

\noindent
{\bf Algorithm A.}

\begin{description}
\item[Step 1.]  Draw random variates $u_1, u_2, u_3\sim {\rm uniform(0,1)}$ independently.  Compute $X=-\log(u_1),$\ 
$Y=-\log(u_2)$.
\item[Step 2.] If $u_3\leq e^{-\delta XY}$ return $(X/\beta,\ Y/\gamma)$; otherwise go to Step 1.
\end{description}

For a general rejection algorithm, its acceptance rate is the ratio of the area under $f(x, y)$ over that under the envelope 
function, which for algorithm A simplifies to 
$$R_A(\delta)=\frac{\int\int f(x, y) dx dy}{\int\int g_0(x, y) dx dy}=\int_0^\infty e^{-x}(1+\delta x)^{-1} dx.$$
It is easy to show that $R_A(\delta)\rightarrow 1$ as $\delta\rightarrow 0$ and $R_A(\delta)\rightarrow 0$ as 
$\delta\rightarrow \infty$.  In other words, for large $\delta$, the expected number of trials until a pair $(X, Y)$ is 
accepted can be unreasonably high.  

An alternative strategy is to first sample $X$ according to its marginal density, and then sample $Y$ given $X$.  Let us 
assume for notational convenience $\beta=1$ (a simple scaling gives the corresponding result for general $\beta$).  The 
(unnormalized) marginal density of $X$ is given by
\begin{equation}
f_X(x)=e^{-x}(1+\delta x)^{-1},
\label{densX}
\end{equation}
and the conditional of $Y$ given $X$ is 
$$f_{Y|X}(y|x)\propto e^{-(1+\delta x)\gamma y}.$$
That is, $Y|X \sim {\rm exponential(1)}/[\gamma(1+\delta X)]$.  To accomplish the more difficult part of sampling $X$ 
according to (\ref{densX}), consider the function
$$g(x;\, c)=\Bigg\{\begin{array}{ll} (1+\delta x)^{-1} & 0<x< c\\ e^{-x} (1+\delta c)^{-1} & x\geq c \end{array},$$
where $c\geq 0$ is a constant to be determined.  Clearly 
$$f_X(x)\leq g(x;\, c),\quad x>0,$$ 
hence $g(x;\, c)$ is a legitimate envelope function for all $c\geq 0$.  Drawing $X$ according to $g(x;\, c)$ is simple, 
because $g(x;\, c)$ is a mixture whose two components are both easy to sample via inversion.  Specifically
$$g(x;\, c)= d_1g_1(x;\, c)+ d_2g_2(x;\, c),$$
where 
$$\begin{array}{lll}
d_1=\delta^{-1}\log(1+\delta c),& g_1(x;\, c)=\delta/[(1+\delta x)\log(1+\delta c)],\ &0<x<c;\\
d_2=e^{-c}/(1+\delta c),& g_2(x;\, c)=e^{-x+c},\ &x\geq c.
\end{array}$$

Both $g_1(x;\, c)$ and $g_2(x;\, c)$ are normalized densities.  If we draw $X$ according to $g_1$ with probability 
$d_1/(d_1+d_2)$, and according to $g_2$ with the remaining probability, then $X$ is distributed according to $g$ overall.  
This yields the following algorithm for sampling from the original bivariate density.

\noindent
{\bf Algorithm B.}
\begin{description}
\item[Step 0.]  Compute $d_1=\delta^{-1}\log(1+\delta c)$ and $d_2=e^{-c}/(1+\delta c)$.
\item[Step 1.]  Draw random variates $u_0,\ u_1,\ u_2\sim {\rm uniform(0, 1)}$ independently.  
\item[Step 2.]  When $u_0< d_1/(d_1+d_2)$, set $X=((1+c\delta)^{u_1}-1)/\delta$; if $u_2<e^{-X}$ go to Step 3, otherwise go 
to Step 1.  When $u_0>d_1/(d_1+d_2)$, set $X=c-\log(u_1)$; if $u_2<(1+\delta c)/(1+\delta X)$ go to Step 3, otherwise go to 
Step 1.
\item[Step 3.]  Draw $u_3\sim {\rm uniform(0, 1)}$.  Return $(X/\beta,\ -\log(u_3)/[\gamma (1+\delta X)] )$.
\end{description}

Note that $d_1$ and $d_2$ may be pre-computed if many random variates with the same parameter $\delta$ are desired.  The 
acceptance rate of algorithm B is easily obtained as
$$R_B(\delta;\, c)=\frac{\int f_X(x) dx}{\int g(x;\, c) dx}=\frac{\int_0^\infty e^{-x}(1+\delta x)^{-1} dx}{d_1+d_2}.$$
A natural question is how to determine $c$.  If $\delta$ is small, say $\delta<1$, choosing $c=0$, which amounts to using an 
exponential envelope, results in a reasonable acceptance rate.  Algorithm B reduces to Algorithm C in this case.

\noindent
{\bf Algorithm C.}
\begin{description}
\item[Step 1.]  Draw random variates $u_1,\ u_2\sim {\rm uniform(0, 1)}$ independently.
\item[Step 2.]  Set $X=-\log(u_1)$; if $u_2<(1+\delta X)^{-1}$ go to Step 3, otherwise go to Step 1.
\item[Step 3.]  Draw $u_3\sim {\rm uniform(0, 1)}$.  Return $(X/\beta,\ -\log(u_3)/[\gamma (1+\delta X)] )$.
\end{description}

The acceptance rate of Algorithm C is 
$$R_C(\delta)=R_B(\delta;\, 0)=\int_0^\infty e^{-x}(1+\delta x)^{-1} dx,$$
which coincides with that of Algorithm A.  Algorithm C is therefore unsuitable for large $\delta$.
(Note that, given their identical acceptance rates, Algorithm C has a slight advantage over Algorithm A because Algorithm C
uses two uniform variates whereas Algorithm A uses three for every rejected sample.)

In contrast, the following shows, for each $c>0$, a positive lower bound of the acceptance rate of Algorithm B over the range 
of $\delta$.

\begin{proposition}
If $\delta>0$ and $c>0$ then
\begin{equation}
R_B(\delta;\, c)\geq (e^c+c^{-1})^{-1}.
\end{equation}
\end{proposition} 

{\it Proof.} We have
\begin{align*}
R_B(\delta;\, c)&=(d_1+d_2)^{-1}\int_0^\infty e^{-x}(1+\delta x)^{-1} dx\\
                &\geq (d_1+d_2)^{-1}\int_0^c e^{-c}(1+\delta x)^{-1} dx\\
                &=e^{-c}d_1(d_1+d_2)^{-1}.
\end{align*}
But $d_1/d_2=e^c \delta^{-1}(1+\delta c)\log(1+\delta c)\geq ce^c$, where we have used a simple inequality:
$(1+x)\log(1+x)\geq x$ when $x\geq 0$.  Thus
\begin{align*}
R_B(\delta;\, c)&\geq e^{-c}d_1(d_1+d_2)^{-1}\\
                &\geq e^{-c}ce^c(ce^c+1)^{-1}\\
                &=(e^c+c^{-1})^{-1}. \qed
\end{align*}
For large $\delta$ we need Algorithm B with a good choice of $c$.  Though it is desirable to choose $c$ such that 
 $R_B(\delta;\, c)$ is optimized, this optimization is difficult analytically.  Time consumed to locate the exact maximizer 
of $R_B(\delta;\, c)$ may well offset the improved acceptance rate, especially if $\delta$ changes frequently.  Fortunately, 
it is observed that, when $0.5<c<1$, $R_B(\delta;\, c)$ is quite insensitive to the value of $c$ over the full range of 
$\delta$.  Table 1 gives the acceptance rate $R_B(\delta;\, c)$ for various values of $\delta$ and $c=0,\ 0.5,\ 0.7,\ 1$.  
Note that Algorithm B does not apply if $\delta=0$.  The column for $\delta=0$ is taken as $\lim_{\delta\rightarrow 0} 
R_B(\delta;\, c)$.

\begin{table}
\caption{Acceptance rate $R_B(\delta;\, c)$ of Algorithm B for various values of $\delta$ and $c$.}
\begin{center}
\begin{tabular}{l|cccccccccccc}
\hline\ & \multicolumn{12}{c}{$\delta$}\\
$c$    & 0    & 0.1  & 0.2  & 0.5  & 1    & 1.5  & 2    & 3    & 5    & 10   & 20   & 100\\
\hline
$0$    & 1.00 & .916 & .852 & .723 & .596 & .517 & .461 & .386 & .299 & .201 & .130 & .041\\
$0.5$  & .904 & .859 & .829 & .776 & .736 & .719 & .710 & .704 & .705 & .719 & .741 & .796\\
$0.7$  & .836 & .803 & .781 & .747 & .725 & .718 & .716 & .718 & .726 & .746 & .770 & .822\\
$1$    & .731 & .711 & .700 & .684 & .680 & .682 & .687 & .696 & .712 & .737 & .764 & .819\\
\hline
\end{tabular}
\end{center}
\end{table}

With $c=0.7$, $R_B(\delta;\, c)$ has an approximate lower bound of $0.716$.  On the other hand, for $\delta<1$, the 
acceptance rate of Algorithm C, $R_C(\delta)=R_B(\delta;\, 0)$, is bounded below by $0.596$.  Algorithm C has the advantage 
of simplicity.  In addition it avoids the numerical problems of Algorithm B when $\delta$ is near zero.  We recommend 
Algorithm C when $\delta<1$ and Algorithm B with $c=0.7$ otherwise. 

\section*{Acknowledgement}
The author thanks an associate editor and two referees for their valuable comments.


\begin{thebibliography}{99}
\bibitem{AS}
Arnold, B.C., Strauss, D., 1988. Bivariate distributions with exponential conditionals. J. Amer. Statist. Assoc. 83, 522--527.
\bibitem{AS1}
Arnold, B.C., Strauss, D., 1991. Bivariate distributions with conditionals in prescribed exponential families.
J. Roy. Statist. Soc. B, 53, 365--375.
\bibitem{D}
Devroye, L., 1986. Non-Uniform Random Variate Generation, Springer-Verlag, New York.
\bibitem{N}
Nadarajah, S., Kotz, S., 2006. Reliability for some bivariate exponential distributions. "Math. Prob. Eng., vol. 2006.
\end{thebibliography}
\end{document}